\documentclass[12pt]{article}
\usepackage{epsfig}
\usepackage{rotating}
\newcommand{\gsim}{\lower.7ex\hbox{$\;\stackrel{\textstyle>}{\sim}\;$}}
\newcommand{\lsim}{\lower.7ex\hbox{$\;\stackrel{\textstyle<}{\sim}\;$}}
\newcommand{\bm}[1]{{\mbox{\boldmath $#1$}}}
\renewcommand{\AA}{{\bf A}\kern-1.5mm\raisebox{.7mm}{$\scriptstyle
\backslash$}}
\newcommand{\tc}{t_{\rm c}}
\def\sidno{\ifcase\arabic{page}\or
  1\or 2\or 3\or 4\or \or 6\or 7\or 8\or 9\fi }

\begin{document}
\footnotesep=14pt
\begin{flushright}
\baselineskip=14pt
{\normalsize DAMTP-1999-180}\\
{\normalsize {astro-ph/0002307}}
\end{flushright}

\vspace*{.5cm}
\renewcommand{\thefootnote}{\fnsymbol{footnote}}
\setcounter{footnote}{0}
\begin{center}
{\Large\bf The origin of cosmic magnetic fields\footnote{Talk presented
at the Third International Conference on Particle Physics and the
Early Universe (COSMO-99), Trieste, Italy, 27 Sept - 3 Oct, 1999, to
appear in the proceedings.}}
\end{center}
\setcounter{footnote}{2}
\begin{center}
\baselineskip=16pt
{\bf Ola T\"{o}rnkvist}\footnote{E-mail:
{\tt o.tornkvist@damtp.cam.ac.uk}}\\
\vspace{0.4cm}
{\em Department of Applied Mathematics and Theoretical Physics,}\\
{\em Centre for Mathematical Sciences, University of Cambridge,}\\
{\em Wilberforce Road, Cambridge CB3 0WA, United Kingdom}\\
\vspace*{0.25cm}{31 January 2000}
\end{center}
\baselineskip=20pt
\vspace*{.5cm}
\begin{quote}
\begin{center}
{\bf\large Abstract}
\end{center}
\vspace{0.2cm}
{\baselineskip=10pt%
In this talk,
I review a number of particle-physics models that lead to
the creation of magnetic fields in the early universe and address the
complex problem of evolving such primordial magnetic fields into
the fields observed today. Implications for future observations of the
Cosmic Microwave Background (CMB) are discussed. Focussing
on first-order phase transitions in the early universe, I describe
how magnetic fields arise in the collision of expanding true-vacuum
bubbles both in Abelian and non-Abelian gauge theories.}
\end{quote}
\renewcommand{\thefootnote}{\arabic{footnote}}
\setcounter{footnote}{0}
\newpage
\baselineskip=16pt
\section{Introduction to Cosmic Magnetic Fields}

A large number of spiral galaxies, including the Milky Way, carry
magnetic fields \cite{Kronberg}.
With few exceptions,
the galactic field strengths are measured to be
a few times $10^{-6}$ G. This particular value
has also been found at a redshift of
$z=0.395$ \cite{redshift}
and between the galaxies in clusters.

Studies of the polarisation of synchrotron radiation emitted by
galaxies with a face-on view, such as M51, reveal that their magnetic
fields are aligned
with spiral arms and density waves in the disk.
A plausible explanation is that
galactic magnetic
fields were
created by a mean-field dynamo mechanism \cite{dynamo}, in which a
much smaller seed
field was exponentially amplified by the turbulent motion of ionised gas
in conjunction with the differential rotation of the galaxy.

For the
dynamo to work, the initial seed
field must be correlated on a scale of $100$ pc, corresponding to the
largest turbulent eddy \cite{dynamo}.
The required strength of the seed field is subject to large uncertainties;
past authors have quoted
$10^{-21\pm 2}$ G as the lower bound
at the time of completed galaxy formation.
This would present a problem for most particle-physics and field-theory
inspired mechanisms of magnetic field generation. However, in recent
work with A.-C.\ Davis and M.J.\ Lilley \cite{bbounds}, I have shown that
the lower bound on the dynamo seed field can be significantly
relaxed if the universe is flat with a cosmological constant, as
is suggested by recent supernovae observations \cite{supernovae}.
In particular, for
the same dynamo parameters that give a lower bound of
$10^{-20}$ G for $\Omega_0=1$, $\Omega_\Lambda=0$,
we obtain
$10^{-30}$ G for $\Omega_0=0.2=1-\Omega_\Lambda$, implying that
particle-physics mechanisms could still be viable.
The observation at redshift $z=0.395$~\cite{redshift} can also be
accounted for with these parameters, but
requires a seed field of at least $10^{-23}$ G \cite{bbounds}.

The dynamo amplifies the magnetic field until
its energy
reaches equipartition with the kinetic energy of
the
ionised gas, $\langle B^2/2\rangle=\langle\rho v^2/2\rangle$, when further
growth is suppressed by dynamical back reaction.
Thus
a
final field of
$B_0\approx 10^{-6}$ G
results
for any seed field of sufficient strength.

In order to explain the galactic field strength without a
dynamo mechanism,
one would require a strong primordial field of
$10^{-3} (\Omega_0 h^2)^{1/3}$ G  at the epoch of radiation
decoupling $t_{\rm dec}$, corresponding to a
field
strength $10^{-9} (\Omega_0 h^2)^{1/3}$ G
on comoving scales of $1$ Mpc. Future precision measurements
of the CMB will put
severe constraints on such a primordial field \cite{CMB}.
Moreover,
magnetic fields on Mpc scales
have been probed by observations
of the Faraday rotation of polarised light from distant luminous
sources, which give an upper bound of about $10^{-9}$ G \cite{Kronberg}.
The observation of micro-Gauss fields between galaxies
in clusters presents an interesting dilemma. Because such regions are
considerably less dense than galaxies, it is doubtful whether a dynamo
could have been operative. Thus the intra-cluster magnetic fields, unless
somehow ejected from
galaxies, have formed directly from a primordial field stronger
than $10^{-3} (\Omega_0 h^2)^{1/3}$ G at $t_{\rm dec}$. Such a field would
certainly leave a signature in future CMB data \cite{CMB}.

Particle-physics
inspired models, which typically produce weak
seed fields,
lead to precise predictions and
there
have an advantage over astrophysical mechanisms,
where the magnetic field strength is determined by complicated
nonlinear dynamics,
or solutions of general relativity with a magnetic field \cite{Thorne},
where the field strength must be fixed by observations.
With the possible exception of the last-mentioned model,
there is no compelling scenario that produces a
primordial
field strong enough to eliminate the need for
a dynamo.

Seed fields for the dynamo can be astrophysical or primordial.
In the former category there is
the important possibility that a seed field may arise spontaneously
due to non-parallel gradients
of pressure and charge density
during the collapse of a protogalaxy \cite{Kulsrud}. For the rest of
this talk, however, I shall assume that the seed field is primordial.

\section{Primordial Seed Fields}

It is
useful to
distinguish between primordial seed fields that are produced
with correlation
length smaller than
vs.\ larger than the horizon size.

{\sl Subhorizon-scale seed fields} typically arise
in first-order phase transitions and from causal processes involving
defects. For example, magnetic fields may be created on the surface
of bubble walls \cite{Hogan} due to local charge separation
induced by baryon-number gradients. The magnetic fields are then
amplified by plasma turbulence near the bubble wall. This possibility has been
explored for the QCD \cite{Cheng} as well as for the
electroweak \cite{baym-sigl}
phase transition.

The production of magnetic fields in
collisions of expanding true-vacuum
bubbles will be discussed
in Sec.~\ref{bubs}. Fields can also be generated
in the wakes of, or due to the wiggles of, GUT-scale cosmic strings during
structure formation, resulting in a large correlation
length \cite{supercond}. Joyce and Shaposhnikov have shown that an
asymmetry of right-handed electrons, possibly generated at the GUT scale,
would become unstable to the generation of a hypercharge magnetic field
shortly before the electroweak phase transition \cite{Joyce}, leading to a
correlation length of order $10^{6}/T$.

{\sl Horizon-scale seed fields} emerge naturally in second-order phase
transitions of gauge theories
from the failure of
covariant derivatives of the Higgs field to correlate on superhorizon
scales \cite{Vacha}.

{\sl Superhorizon-scale seed fields} can arise as a solution of
the Einstein equations for axisymmetric universes \cite{Thorne} and
in inflationary or pre-Big Bang (superstring) scenarios. In the latter case,
vacuum fluctuations of the field tensor are amplified by the
dynamical
dilaton field \cite{Gasperini}. Inflationary models produce extremely
weak magnetic fields unless conformal invariance is explicitly
broken \cite{TurWid}, but even then great difficulties remain.
An exciting new possibility is that magnetic fields may be produced
via parametric resonance with an oscillating field \cite{Finelli}
e.g.\ during
preheating after inflation.
Because the inflaton is initially coherent on
superhorizon scales, large correlations can arise without violating
causality.
A similar proposal involves
charged scalar particles, minimally coupled to
gravity, that are created from the vacuum due to the changing
space-time geometry at the end of inflation. The particles give
rise to fluctuating electric currents which are claimed to produce
superhorizon-scale (indeed, galactic-scale) fields of sufficient
strength to satisfy the galactic dynamo bound \cite{Calzetta}.
This mechanism deserves further investigation.

\section{Evolution of Primordial Magnetic Fields}

A serious problem with many
particle-physics and field-theory scenarios for producing primordial
magnetic fields is that the resulting correlation length $\xi$ is
very small.
If the fields are produced
at the QCD phase transition or earlier with sub-horizon correlation
length, then the expansion
of the universe cannot stretch
$\xi$ to more than 1 pc today (see Fig.~1).
This is far short of the
galactic dynamo lower bound of 5-10 kpc (comoving), corresponding
to 100 pc in a virialised galaxy \cite{bbounds}.

\begin{sidewaysfigure}
\epsfig{figure=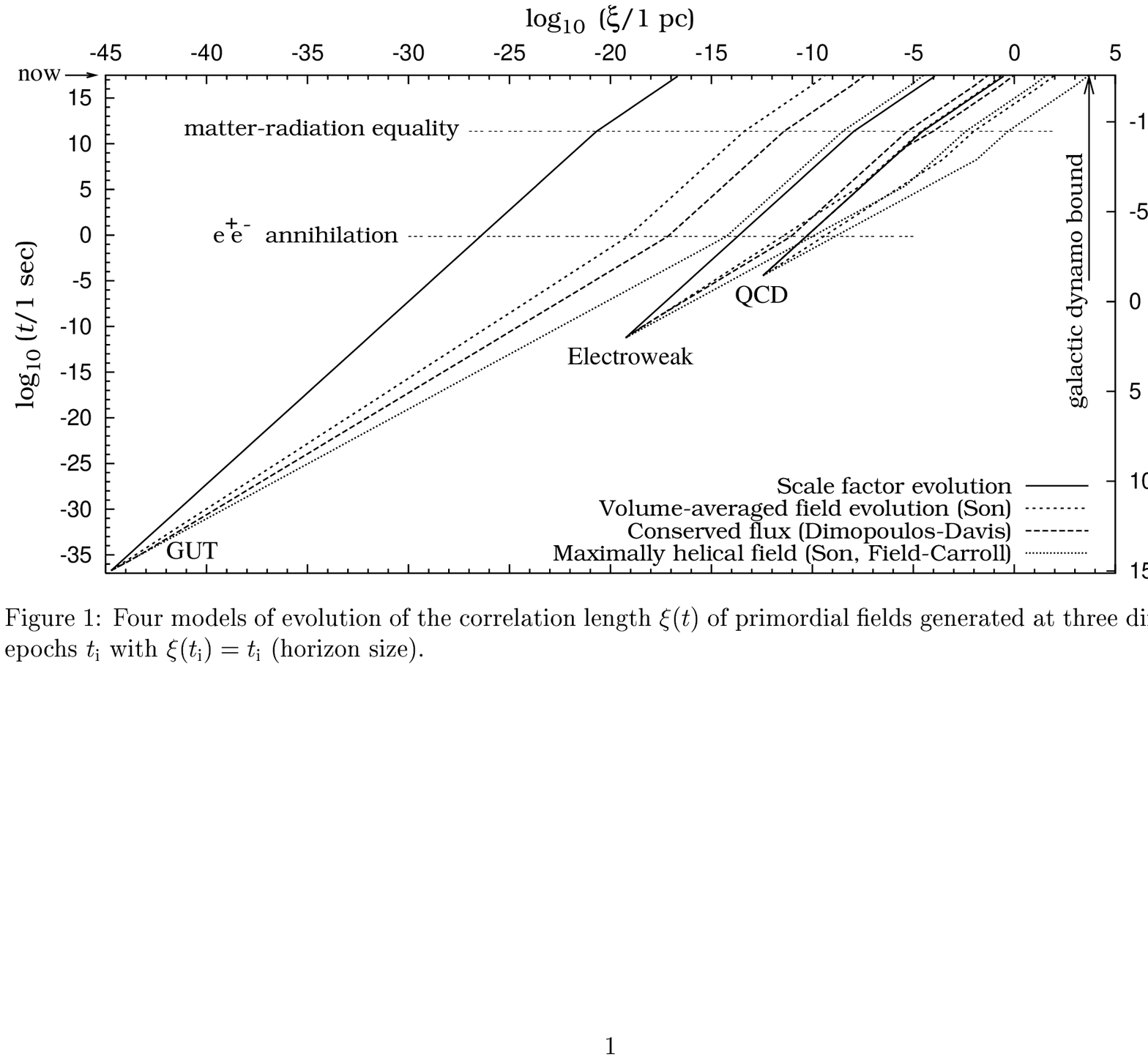,width=\textheight}
\end{sidewaysfigure}

Nevertheless, many authors \cite{brandenburg,Son,carroll,kostas}
have argued that the correlation length
will grow more rapidly due to magnetohydrodynamic (MHD) turbulence
and inverse cascade, which transfers power from
small-scale to large-scale Fourier modes. Several non-relativistic
models for this
evolution are analysed in Fig.~1. The most conservative estimate is
obtained by assuming that
the magnetic field strength on the scale of one correlation
length at any time
equals the volume average of fields that were produced on smaller
scales but have
since decayed \cite{Son}.
This leads to a growth $\xi\sim t^{7/10}$ (obtained from
the Minkowski-space growth $\xi\sim t^{2/5}$ via the substitution
$t\to \tau=t^{1/2}$ and multiplication by the scale factor). The most
optimistic estimate corresponds to the case
when the magnetic field has maximal
helicity in relation to the energy density \cite{Son,carroll}. As magnetic
helicity is approximately conserved
in the high-conductivity early-universe environment, one obtains
the growth law $\xi\sim t^{5/6}$.
Turbulence ends, freezing the growth (in comoving coordinates)
when the kinetic Reynolds number drops
below unity
at the $e^+e^-$ annihilation or later, depending on the model and
the parameters of the initial field.

An intermediate and rather plausible estimate has been given by
Dimopoulos and Davis \cite{kostas}, who use the fact that
the magnetic flux enclosed by a (sufficiently large) comoving closed
curve is conserved. The correlation length here increases at a rate
given by the Alfv\'{e}n velocity, so that $\xi\sim t^{3/4}$.

As Fig.~1 shows, only the most optimistic of these growth laws
leads to a correlation length today
that satisfies the galactic dynamo bound. This occurs
 for fields correlated over the horizon scale at the QCD phase transition.
Beware, however, that the growth laws were derived
using {\sl non-relativistic\/} MHD equations assuming
that the magnetic field energy density remains
in equipartition with the
kinetic energy density
$\rho\bar{v}^2/2$, where $\bar{v}$ is the presumed {\sl non-relativistic\/}
``bulk velocity'' of
the {\sl ultra-relativistic\/} plasma.
It seems plausible that a relativistic treatment could alter the predicted
evolution dramatically.
In this light, I
find it too early to reject the idea that also subhorizon fields
might evolve into fields sufficiently correlated to seed the galactic
dynamo.

At the same time, Fig.~1 demonstrates
the intrinsic advantage of
superhorizon field generation mechanisms.
For these,
the principal problem is not the correlation length,
 but to achieve sufficient strength of the magnetic field.

\section{Magnetic Fields From Bubble Collisions}
\label{bubs}

First-order phase transitions in the early
universe proceed through the nucleation of bubbles \cite{Coleman},
which subsequently expand and collide.
In order to study the generation of magnetic fields,
the initial field strength
is assumed to vanish.
One may then choose
a gauge in which the vector potential $V_\mu$ is initially zero.
In this gauge, the nucleation probability is peaked around bubbles
with constant orientation (phase) of the Higgs field.

We consider first a U(1) toy model.
Let the Higgs field in two colliding bubbles be given by
$\phi_1=\rho_1(x) e^{i\theta_1}$ and
$\phi_2=\rho_2(x) e^{i\theta_2}$, respectively,
where $\theta_1\neq\theta_2$.
When the bubbles meet, the phase gradient establishes a
gauge-invariant current $j_k=iq[\phi^{\dagger} D_k \phi -
(D_k\phi)^\dagger \phi]$ across the surface of intersection of the
two bubbles, where $D_k=\partial_k-iqV_k$. This current, in turn,
gives rise to
a ring-like flux of the field strength $F_{ij}=\partial_i V_j-
\partial_j V_i$, which takes the shape of a girdle encircling the
bubble intersection region.

In recent work we have obtained accurate,
but rather complicated,
 analytical solutions for the
field evolution
in U(1) bubble collisions \cite{u1bub}
 using an analytical expression for the
bubble-wall ``bounce'' profile \cite{Coleman}.
A simpler analytical solution was found by Kibble and
Vilenkin [KV] \cite{KibVil}, who
made
three rather crude
approximations: (1)
The bubble walls move through the plasma without friction,
(2) the modulus of the Higgs field in the
interior of the bubbles
equals a constant, and (3) the phase $\theta$ of the Higgs field is
a step function at the moment of collision. The first of these
assumptions leads to a simple equation of motion for the bubble wall,
which endows the system
with a dynamical O(1,2) symmetry \cite{Hawking}.
All quantities are then functions only of two coordinates:
$z$, the position along the axis through the bubble centres, and
$\tau=\sqrt{t^2-x^2-y^2}$, combining time with the perpendicular directions.
The solutions can be written
down explicitly in terms of Bessel functions
and,
despite the crudeness of the approximations,
capture correctly the qualitative behaviour
of the fields in the bubble overlap
region \cite{u1bub}.

The simplicity of the KV approach makes it ideal for
attacking the more complicated problem of non-Abelian bubble collisions.
These could occur in a first-order electroweak phase transition
(e.g.\ in the MSSM for $M_h\lsim 116$ GeV) or in a GUT phase transition.
Field strengths created in an early phase transition naturally project
onto the electromagnetic U(1) subgroup at the electroweak transition.
I will here concentrate on \mbox{SU(2)$\times$U(1) $\to$U(1)$_{\rm EM}$}
and the minimal Standard Model as a
solvable example.

The initial Higgs field
in the two bubbles can be written in the form
\begin{equation}
\Phi_1=\exp(-i\theta_0\bm{n}\cdot\bm{\sigma})
\left(\begin{array}{c}0\\*\rho_1(x)\end{array}\right)~,\quad\quad
\Phi_2=\exp(i\theta_0\bm{n}\cdot\bm{\sigma})
\left(\begin{array}{c}0\\*\rho_2(x)\end{array}\right)~.
\end{equation}
As the bubbles collide, non-Abelian currents
$j_k^A=i[\Phi^\dagger T^A D_k\Phi - (D_k\Phi)^\dagger T^A \Phi]$
develop across the surface of their intersection, where
$T^A=(g'/2,g\sigma^a/2)$, $D_k=\partial_k-iW_k^A T^A$ and
$W_k^A=(Y_k,W_k^a)$. In analogy with the U(1) case, one obtains
a ring-like flux of non-Abelian fields. The recipe for projecting
out the electromagnetic field
amongst the non-Abelian fields in an
arbitrary gauge has been given elsewhere \cite{bdef}.

In the special cases $\bm{n}=(0,0,\pm 1)$ and $\bm{n}=(n_1,n_2,0)$
it is known \cite{SafCop} that the non-Abelian flux consists of
pure Z and W vector fields, respectively.
The absence of an electromagnetic field
has its explanation in
the fact that the normalised Higgs
field $\hat{\Phi}\equiv
\Phi/(\Phi^\dagger\Phi)^{1/2}$ maps to a geodesic on the
Higgs vacuum manifold and the gauge fields map to a line
in the Lie algebra spanned by the generator of that same geodesic:
\begin{equation}
\hat{\Phi}(x)=\exp(i\theta(x)\bm{n}\cdot\bm{\sigma})
(0~~1)^\top
{}~,\quad
\AA_\mu(x)\equiv W_\mu^A(x) T^A= f_\mu(x)\bm{n}\cdot\bm{\sigma}~.
\end{equation}
When $n_3\neq 0,\pm1$, both $Z$ and $W$ fields are excited. Because
they have unequal masses
$M_W\neq M_Z$, they evolve
differently \cite{Grasso} and the fields $\hat{\Phi}$ and
$\AA_\mu$ stray from the geodesic and its tangent, producing
an electromagnetic current.

I have used a KV approach to derive a
perturbative analytical
solution for the evolution of gauge fields in an electroweak
bubble collision, valid as long as the fields are small and higher-order
nonlinearities can be neglected. The space allotted here
allows me only to indicate the structure of the solution for the
electromagnetic field, which is
\begin{eqnarray}
F^{\alpha z}&=& x^\alpha \frac{\sin\theta_{\rm w}}{g} n_3(1-n_3^2)
\int_0^z dz' \left(\rule{0pt}{5mm}
 h_3(\tau,z')\right.\nonumber\\*
 &+& \int_{\tc}^\tau \!\!d\tau''\!\!
\int_{z'-\tau''+\tc}^{z'+\tau''-\tc} \hspace*{-2mm}
dz'' h_1(\sqrt{(\tau-\tau'')^2 -
(z'-z'')^2}) h_2(\tau'',z'')\!\!\left.\rule{0pt}{5mm}\right),
\end{eqnarray}
where $\tc$ is the time of collision and the functions $h_i$ contain
products of $i$ Bessel functions. As expected, the
resulting field strength is
of the order of $M_W^2/g$ with a correlation length $\xi\sim M_W^{-1}$.
However, when
plasma friction and conductivity are taken into account, the magnetic
field spreads over the interior of a bubble \cite{lilley}
leading to an appreciable increase in correlation length.

\subsection*{Acknowledgments}
The author is supported by the European Commission's TMR programme under
Contract No.~ERBFMBI-CT97-2697.

\end{document}